\documentclass[journal]{IEEEtran}

\usepackage{hhline}
\usepackage{bm}
\usepackage{color}
\usepackage{longtable}	
\usepackage{multicol}
\usepackage{float}
\usepackage{multirow}
\usepackage{cite}
\usepackage{amsmath,amssymb,amsfonts}
\usepackage{algorithmic}
\usepackage{graphicx}
\usepackage{textcomp}

\usepackage{diagbox}
\allowdisplaybreaks
\ifCLASSOPTIONcompsoc
\usepackage[caption=false,font=normalsize,labelfon
t=sf,textfont=sf]{subfig}
\else
\usepackage[caption=false,font=footnotesize]{subfi
g}
\usepackage[dvipsnames]{xcolor}

\IEEEoverridecommandlockouts

\definecolor{BrickRed}{rgb}{0.0 ,0,0}
\definecolor{amethyst}{rgb}{0.6, 0.4, 0.8}
\usepackage{xcolor}

\newcommand{\E}{\mathbb{E}}

\newcommand{\Var}{\mathrm{Var}}

\begin{document}

\title{Analytical Study of the Impacts of Stochastic Load Fluctuation on the Dynamic Voltage Stability Margin Using Bifurcation Theory}


\author{Georgia~Pierrou,~\IEEEmembership{Student Member,~IEEE,} and~Xiaozhe~Wang,~\IEEEmembership{Member,~IEEE}
\thanks{The authors are with the Department of Electrical and Computer Engineering, McGill University, Montr\'{e}al, QC H3A 0G4, Canada. (e-mail: georgia.pierrou@mail.mcgill.ca, xiaozhe.wang2@mcgill.ca)}
\thanks{This work is supported by the Natural Sciences and Engineering Research Council (NSERC) under Discovery Grant NSERC RGPIN-2016-04570, the Fonds de Recherche du Qu\'ebec-Nature et technologies under Grant FRQ-NT PR-253686 and the Stavros S. Niarchos Foundation McGill Fellowship.
}
\thanks{Digital Object Identifier 10.1109/TCSI.2019.2943509}

}

\maketitle

\begin{abstract}
This paper studies the impacts of stochastic load fluctuations, namely the fluctuation intensity and the \textcolor{BrickRed}{load power variation speed}, on power system dynamic voltage stability. Additionally, the trade-off relationship between the two parameters is revealed, which provides important insights regarding the potential of using energy storage to maintain voltage stability under high uncertainty. To this end, Stochastic Differential-Algebraic Equations (SDAEs) are used to model the stochastic load variation; bifurcation analysis is carried out to explain the influence of stochasticity. Numerical study and Monte Carlo simulations on the IEEE 14-bus system demonstrate that a larger fluctuation intensity or a slower load power variation speed may lead to a smaller voltage stability margin. \textcolor{BrickRed}{To the best of authors' knowledge, this work uncovers} the impacts of the time evolution property of the driving parameters, i.e., the \textcolor{BrickRed}{load power variation speed} and its trade off effect with the fluctuation intensity on the size of the dynamic voltage stability margin.
\end{abstract}



\section{Introduction}
\setcounter{page}{1}
Voltage stability analysis is becoming a challenging issue as the load demand continues growing and the integration of \textcolor{BrickRed}{renewable energy sources} increases. While the
fluctuating power output from \textcolor{BrickRed}{renewable energy sources} introduces uncertainty on the generation side, loading patterns entail stochastic fluctuations on the demand side. 

In previous work, 
both static approaches and dynamic approaches have been exploited for power system voltage stability analysis. Traditionally,
voltage collapse was attributed to the loss of power flow feasibility \cite{Galiana84}. Therefore, static approaches such as Continuation Power Flow have been used in \cite{Haesen09} to study the impacts of the variability of \textcolor{BrickRed}{renewable energy sources} and loads on power system voltage stability. \textcolor{BrickRed}{A formula to determine the transmission reliability margin has been developed in \cite{Zhang04}, which accounts for various uncertainties from loads, topology changes, etc.}  

However, dynamic components such as generator and load dynamics play significant roles in voltage stability as shown in \cite{Cutsem}, \cite{Taylor}. \textcolor{BrickRed}{Furthermore, the dynamic evolution of critical parameters, such as the high load variation speed, may result in
catastrophic voltage collapse (see the 1987 Tokyo Blackout \cite{Ohno06}), indicating that a pure static analysis may not be sufficient and the load power variation speed is practically important in voltage stability analysis.}
As a result, a dynamic approach incorporating all dynamic components and control actions should be applied. 
To account for the randomness of \textcolor{BrickRed}{renewable energy sources} and loads in the dynamic stability study, Stochastic Differential-Algebraic Equations (SDAEs) have been proposed in \textcolor{BrickRed}{\cite{Milano13} -- \cite{Nwankpa93}} 
and applied in different power system stability studies \cite{Nwankpa93} --\cite{Li19}. \textcolor{BrickRed}{In \cite{Dhople13}, a stochastic hybrid system is proposed to study the impact of stochastic active and reactive power injections, modeled as continuous Markov chains, on power system dynamic performance.} 

Regarding the dynamic voltage stability, 
an analytical study based on the sample path of SDAEs was conducted in \cite{Wangxz15}, \cite{Wang17} to investigate the impacts of stochastic wind power on dynamic voltage stability, showing that neglecting the variability of wind power may lead to incorrect voltage stability assessment. However, an analytical study close to the bifurcation point is lacking. In \cite{Ghanavati14}, the statistics of the sample paths of the SDAEs models were proposed to predict the voltage collapse of power systems. The mean path of all sample paths was exploited in \cite{Souxes19} to analyze the impacts of wind power variability on the reactive power support control scheme for voltage stability. The authors of \cite{DeMarco87} considered load fluctuations in voltage stability study and leveraged on the first exit time of SDAEs as a measure for voltage stability assessment. Similar approaches were adopted in \cite{Nwankpa93}, \cite{Nwankpa2000} to study the intensities of load fluctuations and the uncertainty of wind power on system voltage stability.

The bifurcation theory for a dynamical system has also been applied to
explain the dynamic mechanisms of voltage collapse \cite{Chiang90} -\cite{Canizares95}, which, nevertheless, did not consider the randomness brought about by the loads and \textcolor{BrickRed}{renewable energy sources}. 
More recently, the authors of \cite{Qiu08} applied the  bifurcation theory and the eigenvalue analysis
to study the effect of stochastic load fluctuations on voltage stability based on a single realization of the stochastic dynamical system.
However, the applied bifurcation theory was for the deterministic rather than the stochastic dynamical system. Also, the implementation of Monte Carlo simulations to statistically describe the impact of uncertainty is lacking. In addition, in all aforementioned studies, the focus was given to the impacts of the stochastic fluctuation strength on the voltage stability. \textcolor{BrickRed}{To the best of authors' knowledge, this work presents for the first time} another important parameter that will also affect the dynamic voltage stability margin.

In this paper, we aim to analytically investigate the impacts of stochastic load fluctuation  
on the dynamic voltage stability margin, leveraging on the bifurcation theory for the stochastic dynamical system. 
Particularly, it will be shown that the fluctuation intensity is not the only factor affecting the size of the voltage stability margin. 
Interestingly, the time evolution property of the driving parameters, e.g, the \textcolor{BrickRed}{load power variation speed}, also plays significant role. The trade-off relationship between the fluctuation intensity and the power variation speed
provides important insights towards the design of control measures to maintain the voltage stability of power systems under high uncertainty. The main contributions of the paper are as follows:

\begin{itemize}

    \item We analytically and numerically show the existence of a strong/weak noise regime in which the uncertainty affects/does not affect the level of the voltage stability margin. Such result yields an important guideline regarding whether and under what conditions the randomness of loads needs to be considered in voltage stability assessment.
    \item The separation of the strong and weak noise regimes depends on two parameters, namely, the load fluctuation intensity and the \textcolor{BrickRed}{load power variation speed}, the impacts of which on the size of the margin are systematically investigated by extensive Monte Carlo simulation.
    \item The trade-off relationship between the load fluctuation intensity and the \textcolor{BrickRed}{load power variation speed} is discussed, which 
    implies a potential of using \textcolor{BrickRed}{energy storage systems} to maintain voltage stability under high uncertainty level.

\end{itemize}

Although it is intuitive that decreasing the variability level helps maintain the voltage stability, the bifurcation theory for the stochastic dynamical system provides essential information regarding how much reduction of the intensity is needed to maintain the same level of the load margin. Such information may help reduce the operating cost of \textcolor{BrickRed}{energy storage systems} while maintaining the stability of the grid with high uncertainty.



 The remainder of the paper is organized as follows: Section \ref{1} introduces the stochastic dynamic power system model in the form of SDAEs.
Section \ref{2} briefly  reviews the bifurcation theory and the theoretical results about the behavior of
the stochastic slow-fast system. 
 Section \ref{3} presents an analytical study of the impacts of stochastic load variation on the voltage stability margin by applying the theoretical results of the stochastic slow-fast systems. 
 Section \ref{4} provides systematic numerical results on the IEEE 14-bus system to validate the analytical results in Section \ref{3}.
 Section \ref{5} presents the conclusions and perspectives.

\section{The Stochastic Power System Model}\label{1}
\subsection{The Stochastic Dynamic Load Model}

The total load seen by a bulk power delivery transformer can be modelled as an aggregated load comprised of many individual physical loads and devices \cite{Cutsem}. In this paper, we consider a generic aggregated model of self-restoring load, which plays significant roles in dynamic voltage stability \cite{Cutsem}, \cite{Hill93}.
The active and reactive consumption of the aggregated load can be described as:
\begin{equation}
\label{eq:erl_power}
\begin{gathered}
p=x_p/T_p+p_t\\
q=x_q/T_q+q_t
\end{gathered}
\end{equation}
where $x_p$ and $x_q$ are the state variables given by:
\begin{equation}
\label{eq:erload}
\begin{gathered}
\dot{x}_p=-x_p/T_p+p_s-p_t \\
\dot{x}_q=-x_q/T_q+q_s-q_t
\end{gathered}
\end{equation}
${T_{p}}$ and ${T_{q}}$ are the corresponding power time constants; ${{p}_{s}}$ and ${{p}_{t}}$ are the static and transient real power absorptions; ${{q}_{s}}$ and ${{q}_{t}}$ are the static and transient reactive power absorptions.

Particularly, $p_s$, $p_t$, $q_s$ and $q_t$ depend on voltages at load buses. Hence, stochastic load variations can be incorporated into the aggregated load model as follows 
\cite{Milano13}, \cite{Wangxz:2017}:
\begin{equation}
\label{eq:exp_rec_load_sde}
\begin{gathered}
{{p}_{s}}={({p}_{0}+\eta_{i}(t))}(\frac{V}{{V}_{0}})^{{\alpha}_{s}} \quad {{p}_{t}}={({p}_{0}+\eta_{i}(t))}(\frac{V}{{V}_{0}})^{{\alpha}_{t}} \\
{{q}_{s}}={({q}_{0}+\eta_{i}(t))}(\frac{V}{{V}_{0}})^{{\beta}_{s}} \quad
{{q}_{t}}={({q}_{0}+\eta_{i}(t))}(\frac{V}{{V}_{0}})^{{\beta}_{t}} \\
\end{gathered}
\end{equation}
where ${{p}_{0}}$ and ${{q}_{0}}$ are the nominal active and reactive power; ${{\alpha}_{s}}$, ${{\beta}_{s}}$, ${{\alpha}_{t}}$ and ${{\beta}_{t}}$ are exponents related to the steady state and the transient load response, respectively; ${{V}_{0}}$ is the nominal bus voltage.  Particularly, $\eta_i$ is a stochastic process describing the stochastic perturbations around the nominal power. It should be noted that
similar procedures can be applied to other dynamic load models such as the frequency-dependent load model, the thermostatic recovery load model, etc., to include the stochastic perturbations into their nominal powers.


In this paper, the stochastic variations $\bm{\eta}$ are modelled as a vector Gauss-Markov process. Although the assumption of white noise may initially not appear obvious, previous works (e.g.,\cite{DeMarco87}) have shown its appropriateness in long-term dynamic voltage stability study, since the stochastic fluctuation is the aggregate behavior of many thousands of individual customer devices switching independently. 
Therefore, we model the load fluctuations by a vector Ornstein-Uhlenbeck  process $\bm{\eta}$, which is Gaussian and Markovian, similar to the approach adopted in \cite{Milano13}, \cite{Wangxz:2017}:
\begin{equation}
\label{eq:ouprocess}
\dot{\bm{\eta}} = -A_{\bm{\eta}}{\bm{\eta}} + \sigma B_{\bm{\eta}}{\bm{\xi}}, \quad t \in [0,T] \end{equation}
where  $A_{\bm{\eta}}=\mbox{diag}([\alpha_{1},...,\alpha_{k}])$ is positive definite and is related with the {correlation time of the load variations \cite{Ghanavati16}}; $\sigma$ describes the intensity of stochastic perturbations;  $B_{\bm{\eta}}=\mbox{diag}([\beta_{1},...,\beta_{k}])$ denotes the relative strength between perturbations; $\bm{\xi}$ is a vector of independent Gaussian random variables, since $\xi_{i}=\frac{d{W}_{ti}}{dt}$ and $W_{ti}$ is a Wiener process.

If the initial condition is $\eta_i(0) \sim \mathcal{N}(0,(\sigma\beta_i)^2/2\alpha_i)$, then the stochastic process $\eta_i$ is a stationary autocorrelated Gaussian process with the following statistical properties \cite{Gardiner}:
\begin{itemize}

\item $\E[\eta_i(t)]=0, \quad  \forall t \in [0,T],$
\item $\Var[\eta_i(t)]=(\sigma\beta_i)^2/2\alpha_i, \quad \forall t \in [0,T],$

\item $\mbox{Aut}[\eta_i(t_{p}),\eta_{i}(t_{q})] = e^{-\alpha_i |t_{q}-t_{p}|}, \quad \forall t_{p},t_{q} \in [0,T]$.

\end{itemize}
\textcolor{BrickRed}{It is worth noting that the parameters $\beta_{i}$ are usually adjusted to result in the desired variance. For instance, a common approach is to choose $\beta_{i}=\sqrt{2\alpha_{i}}$ to remove the influence of $\beta_{i}$ on the variance, so that $Var[\eta_i(t)]=\sigma^2$ \cite{Minano13}. We follow this approach in this work such that $\sigma$ is the sole parameter representing the influence of the variance of $\eta_i$.} 

\vspace{-5pt}
\subsection{SDE Formulation of the Dynamic Power System Model}
With (\ref{eq:erl_power})-
(\ref{eq:ouprocess}), the conventional Differential-Algebraic Equations (DAE)-based power system model incorporating the randomness can be described as:
\begin{equation}
\label{eq:sdae_model}
\begin{gathered}
\dot{\bm{x}} = \bm{\textcolor{BrickRed}{h_1}}(\bm{x},\textcolor{BrickRed}{\bm{z}},\bm{p}, \bm{\eta}) \\
\bm{0}=\textcolor{BrickRed}{\bm{h_2}}(\bm{x},\textcolor{BrickRed}{\bm{z}}, \bm{p})
\end{gathered}
\end{equation}
where $\bm{x}$ is the vector of state variables, e.g., the states of dynamic loads, generator rotor angles;
$\textcolor{BrickRed}{\bm{z}}$ is the vector of algebraic variables, e.g., bus voltage magnitudes and phases;
$\bm{p}$ is the vector of parameters, e.g., load powers;
$\bm{\eta}$ is the vector of stochastic perturbations describing, for instance, load fluctuations or renewable generation variations. $\bm{\textcolor{BrickRed}{h_1}}$ are the differential equations describing the dynamics of system components.
$\textcolor{BrickRed}{\bm{h_2}}$ are the algebraic equations
describing power flow, network connectivity and internal static behaviors of passive devices. $\bm{\textcolor{BrickRed}{h_1}}$  and $\textcolor{BrickRed}{\bm{h_2}}$ are assumed to be sufficiently smooth.

In normal operating conditions, the algebraic Jacobian matrix $\textcolor{BrickRed}{\partial_{\bm{z}}\bm{h_2}}$ is typically non-singular \cite{Bompard96}. Hence, $\textcolor{BrickRed}{\bm{z}}$ can be expressed by $\bm{x}$ and $\bm{p}$, and thus be eliminated according to the Implicit Function Theorem \cite{Cutsem}. As such,  (\ref{eq:ouprocess})-(\ref{eq:sdae_model}) take the following form of Stochastic Differential Equations (SDEs):
\begin{eqnarray}
\label{eq:sde_model_withouty_1}
\dot{\bm{x}} &=& \bm{H}(\bm{x},\bm{p}, \bm{\eta}) \\
\dot{\bm{\eta}} &=& -A_{\bm{\eta}}{\bm{\eta}} +\sigma B_{\bm{\eta}}{\bm{\xi}}
\label{eq:sde_model_withouty_2}
\end{eqnarray}
Denote $\bm{u}=\begin{bmatrix} {\bm{x}}, {\bm{\eta}}
\end{bmatrix}^T$ and $B=\begin{bmatrix} \bm{0}, B_{\bm{\eta}} \end{bmatrix}^T$, then  (\ref{eq:sde_model_withouty_1})-(\ref{eq:sde_model_withouty_2}) can be represented as:
\begin{equation}
\label{eq:sde_model_u}
\dot{\bm{u}} = \bm{G}(\bm{u},\bm{p})+\sigma B \bm{\xi} \\
\end{equation}

\section{Mathematical Preliminaries}
\label{2}
\subsection{The Bifurcation Theory} 
Bifurcation theory has been widely used in the literature to explain 
voltage stability in power systems \cite{Chiang90} -\cite{Canizares95}. Generally, we consider a time-dependent dynamic system with slowly varying parameters described by:
\begin{equation}
\bm{\dot{x}}= \bm{F}(\bm{x},\epsilon t)
\label{general SNB}
\end{equation}
%
where $\bm{x}\in \mathbb{R}^n$ are the state variables.
Passing to the slow time scale $s=\epsilon t$ results in the form of
$\epsilon\bm{{x}^\prime}= \bm{F}(\bm{x},s)$,
which can be further represented as a slow-fast system:
\begin{eqnarray}
\label{eq:ode_slow fast}
\epsilon\bm{{x}^\prime}&=&\bm{F}(\bm{x},y)\nonumber\\
{{y}^\prime}&=&1
\end{eqnarray}
where $^\prime=\frac{d}{ds}$, $\textcolor{BrickRed}{y(s)=s}$.

In power system voltage stability study, slowly varying parameters are typically real and/or reactive power of loads and renewable generators.
%
%
As the parameters vary, a bifurcation may occur leading to a qualitative change in the behavior of the system, such as the change of stability of the equilibrium point (e.g., transcritical bifurcation), the emergence of oscillations (e.g., Hopf bifurcation)
and even the disappearance of equilibrium points (e.g., saddle-node bifurcation). 

\begin{figure}[!b]
 \centering
 \includegraphics[width=2in ,keepaspectratio=true,angle=0]{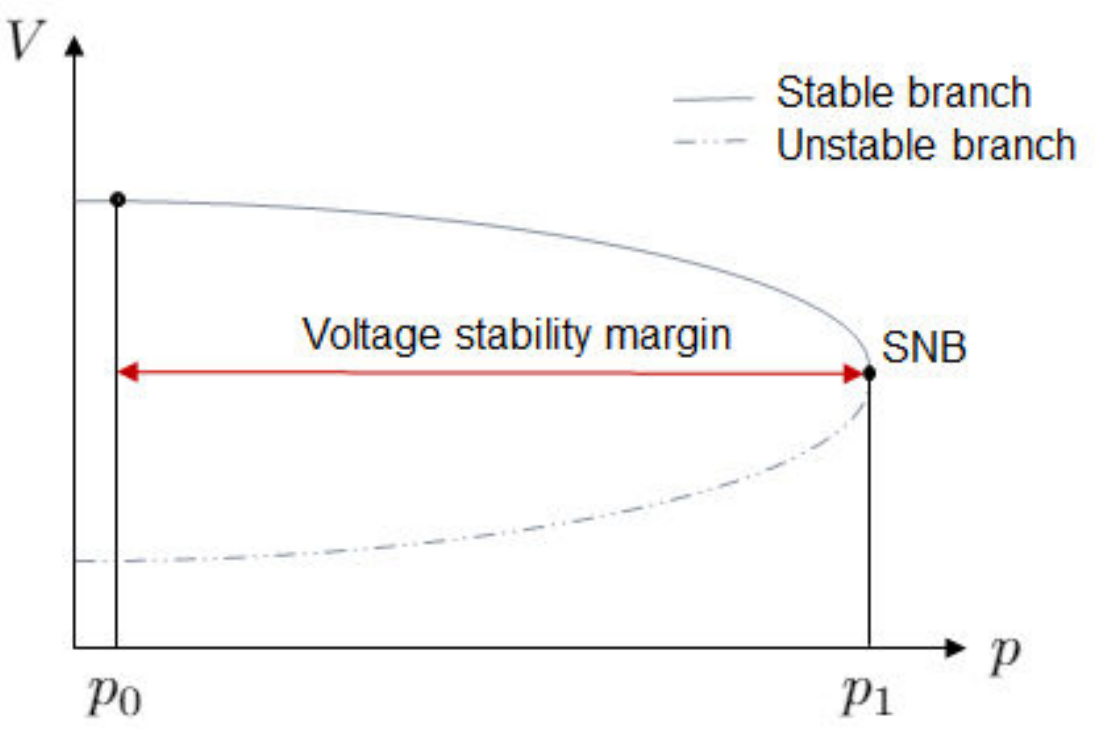}
 \caption{An illustration for the voltage stability margin. 
 }\label{pv_curve}
 \end{figure}

The saddle-node bifurcation (SNB) is typically used to explain the dynamic mechanism of voltage collapse \cite{Chiang89}.
As the parameters $\bm{p}$ (e.g., load powers) change slowly, the equilibrium point $\bm{x^\star}$ will vary in the state space leading to a slow decrease in voltage magnitudes. At the critical load power $\bm{p_1}$, the voltage magnitudes sharply decrease and the system loses stability by $\bm{x^\star}$ disappearing in a SNB.
When a SNB happens, two equilibrium points---one stable and one unstable---will coalesce and disappear. Therefore,
a necessary condition for the SNB is the singularity of the Jacobian matrix $F_{\bm{x}}$ \cite{Canizares92}. The difference between the power at the current operating point and the power at the SNB point is defined as the \textit{voltage stability margin} of the system, as illustrated in Fig. \ref{pv_curve}. The voltage stability margin is commonly used as a stability index to assess the voltage stability of a power system, since it defines the critical value of the parameter (in our case the maximum load power $\bm{p_1}$ following a load increase) for which the system can remain voltage stable \cite{Gao96}.










\subsection{The Slow-Fast System }
\label{slowfastsystem_subsection}


Since the power system model (\ref{general SNB}) can be represented as a slow-fast system (\ref{eq:ode_slow fast}),
we first consider a general slow-fast system:
\begin{eqnarray}
    \epsilon\bm{\dot{x}}&=&\bm{f(x,y)}, \quad \bm{x}\in \mathbb{R}^{n_x}\nonumber\\
    \bm{\dot{y}}&=&\bm{g(x,y)}, \quad \bm{y}\in \mathbb{R}^{n_y}
    \label{general slow fast}
\end{eqnarray}
where $\epsilon$ is a small positive parameter,  \textcolor{BrickRed}{$\bm{f}$ and  $\bm{g}$ sufficiently smooth functions}. $\bm{x}$ is a vector of fast variables while $\bm{y}$ is a vector of slow variables.

Let $\epsilon=0$, the algebraic equation $\bm{0=f(x,y)}$ constraints the slow dynamics to the slow manifold: $\mathcal{M}_{0}$ $=$ \{ $\bm{(x^{\star}(y),y): f(x^{\star}(y)},\bm{y)=0}$, $\bm{y} \in D_{\bm{y}}$  $\subset \mathbb{R}^{n_y}$ \}.  If 
all eigenvalues of the Jacobian matrix $\partial_{\bm{x}} \bm{f(x^\star(y),y)}$ have negative real parts, uniformly bounded away from 0 for $\bm{y}\in D_{\bm{y}}$, then $\mathcal{M}_0$ is \textit{a stable component of the constraint manifold}, i.e., \textit{stable branch}. 

The Fenichel's theorem (\textit{Theorem 1}) states that all trajectories starting near the stable component of the constraint manifold actually converge to an invariant manifold.

\begin{figure}[!tb]
\centering
\includegraphics[width=2.7in ,keepaspectratio=true,angle=0]{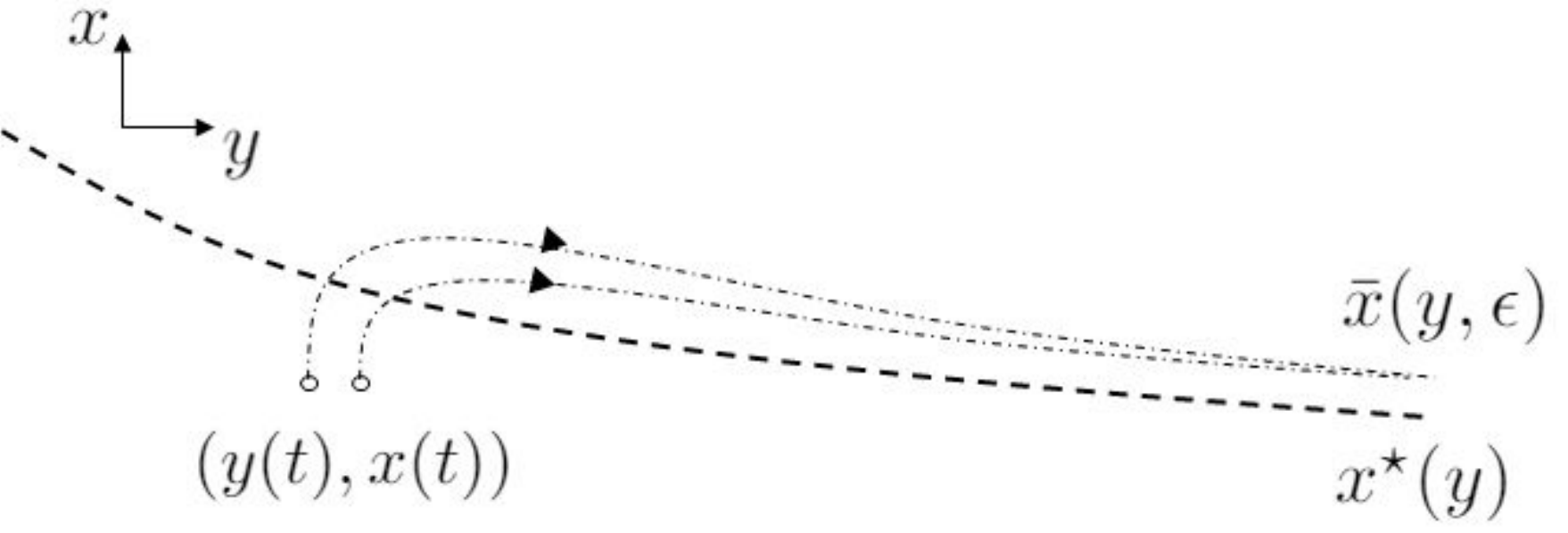}
\caption{Orbits approaching $\mathcal{M}_{\epsilon}$ exponentially fast. }\label{Fenichelstheorem}
\end{figure}

\textit{Theorem 1 (Fenichel 1979 \cite{Fenichel79}):} If $\mathcal{M}_0$ is a stable component of the constraint manifold, then there exists a manifold $\mathcal{M}_{\epsilon}$ $=$ \{ $\bar{\bm{x}}(\bm{y},\epsilon)=\bm{x^\star(y)}+O(\epsilon)\}$, $\epsilon$-close to $\mathcal{M}_0$, that attracts neighboring trajectories exponentially fast, as shown in Fig. \ref{Fenichelstheorem}.

Adding white noise to both the fast and the slow dynamics results in the SDEs shown as below:
\begin{eqnarray}
    \dot{\bm{x}}&=&\frac{1}{\epsilon}\bm{f(x,y)}+\frac{\sigma}{\sqrt{\epsilon}}\bm{f_1(x,y)}\bm{\xi_1}, \quad \bm{x}\in \mathbb{R}^{n_x}\nonumber\\
    \dot{\bm{y}}&=&\bm{g(x,y)}+ \textcolor{BrickRed}{\tilde{\sigma}} \bm{g_1(x,y)}\bm{\xi_2},\qquad \bm{y}\in \mathbb{R}^{n_y} \label{general slow fast SDE}
\end{eqnarray}
where $\bm{\xi_1}$ and $\bm{\xi_2}$ are $k$-dimensional vectors of independent Gaussian random variables respectively, and $\bm{f_1}, \bm{g_1}$
are sufficiently smooth functions. The small parameters $\sigma$ and $\textcolor{BrickRed}{\tilde{\sigma}}$ measure the intensity of the two noise terms. We are interested in the case that $\textcolor{BrickRed}{\tilde{\sigma}}$ does not dominate $\sigma$, i.e.,
$\textcolor{BrickRed}{\tilde{\sigma}}=\rho\sigma$ where $\rho$ is uniformly bounded above in $\epsilon$.

As shown in Appendix \ref{appendix_samplepath}, the noise terms will not greatly affect the sample paths of the stochastic system (\ref{general slow fast SDE}) near the stable component of the constraint manifold compared to the trajectory of the deterministic system (\ref{general slow fast}). However, the impact of stochasticity may be distinct near the bifurcation point, leading to an early transition.  

\subsubsection{Near the Stable Component of the Constraint Manifold}
\begin{figure}[!tb]
\centering
\includegraphics[width=2.5in ,keepaspectratio=true,angle=0]{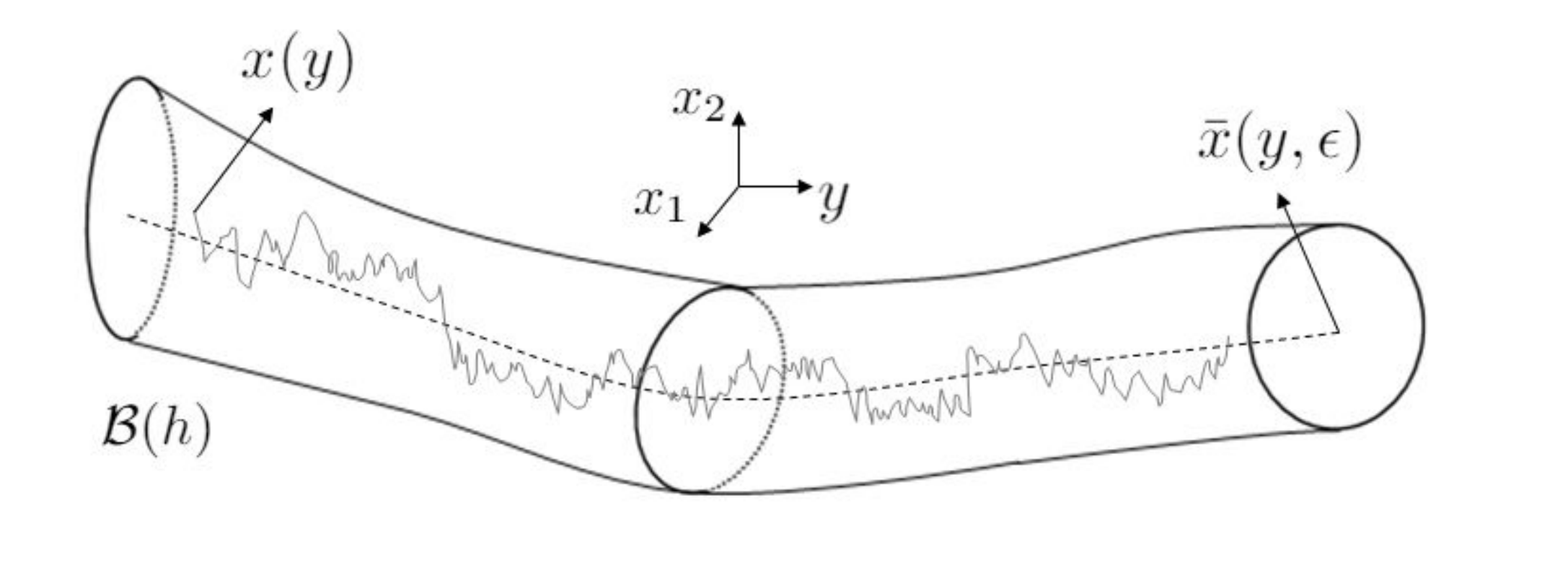}
\vspace{-5pt}
\caption{
The trajectory of the stochastic model (\ref{general slow fast SDE}) is unlikely to leave the neighborhood $\cal{B}$$(h)$ (\ref{neighborhood}) of the trajectory of the deterministic model (\ref{general slow fast}) if  $h\gg\sigma$.}
\label{samplepath}
 \end{figure}

As shown in \textit{Theorem 4} in Appendix \ref{appendix_samplepath} the sample paths of the stochastic system (\ref{general slow fast SDE}) are concentrated in an ellipsoidal layer ${\cal B}(h)$ surrounding the invariant manifold ${\cal M}_{\epsilon}$ of system (\ref{general slow fast}) if $h\gg\sigma$, as illustrated in Fig. \ref{samplepath}.
${\cal B}(h)$ is defined as:
\begin{equation}
{\cal B}(h) = \{( \bm{x,y}):\langle \bm{x}-\bm{\bar{x}}(\bm{y},\epsilon),\bm{\bar{X}}(\bm{y},\epsilon)^{-1}( \bm{x-\bar{x}}(\bm{y},\epsilon)) \rangle<h^2 \}  \label{neighborhood}
\end{equation}
where \textcolor{BrickRed}{$\langle, \rangle$ denotes the inner product}. $\bm{\bar{X}}(
\bm{y},\epsilon)$ describing the shape of the cross section of ${\cal B}(h)$ is well defined as shown in Appendix B in \cite{Wangxz15}. 

This analytical result implies that stochastic perturbation will not greatly affect the trajectory of the slow-fast system near the stable branch.

\subsubsection{Near the Saddle-Node Bifurcation Point}
Consider a particular two-dimensional slow-fast system, i.e., $n_x=1$, $n_y=1$ in (\ref{general slow fast}).
We say that a point $(x^\star, y^\star)$ is the  \textit{saddle-node bifurcation point of system (\ref{general slow fast})} if the fast vector field satisfies the following conditions:
\begin{eqnarray}
&&f(x^\star,y^\star)=0,\qquad
\partial_x f(x^\star,y^\star)=0\\
&&\partial_{xx} f(x^\star,y^\star)\not=0\quad
\partial_{y} f(x^\star,y^\star)\not=0\nonumber
\end{eqnarray}
To simplify analysis, we can always convert the coordinate system and conduct scaling in $x$, $y$ and time to  ensure the following conditions are satisfied at SNB \cite{Gentz06}:
\begin{eqnarray}
&&\mbox{\textit{A1: }} (x^\star,y^\star)=(0,0), \quad
\mbox{\textit{A2: }} \partial_{xx} f(x^\star,y^\star)=-2,\nonumber\\
&&\mbox{\textit{A3: }} \partial_{y} f(x^\star,y^\star)=-1,\quad
\mbox{\textit{A4: }} g(x^\star,y^\star)=1.\nonumber
\end{eqnarray}
\textcolor{BrickRed}{The simplest system that satisfies the aforementioned conditions is shown as below:} 
\begin{eqnarray}
\epsilon\dot{x}&=&-y-x^2\nonumber\\
\dot{y}&=&1\label{normal form near SNB}
\end{eqnarray}
\textcolor{BrickRed}{which is the normal form around SNB for the general slow-fast system (\ref{general slow fast}) \cite{Golubitsky}, \cite{Kuznetsov}.
Note that $(0,0)$ is the SNB point.}

\begin{figure}[!b]
    \centering
    \includegraphics[width=1.5in ,keepaspectratio=true,angle=0]{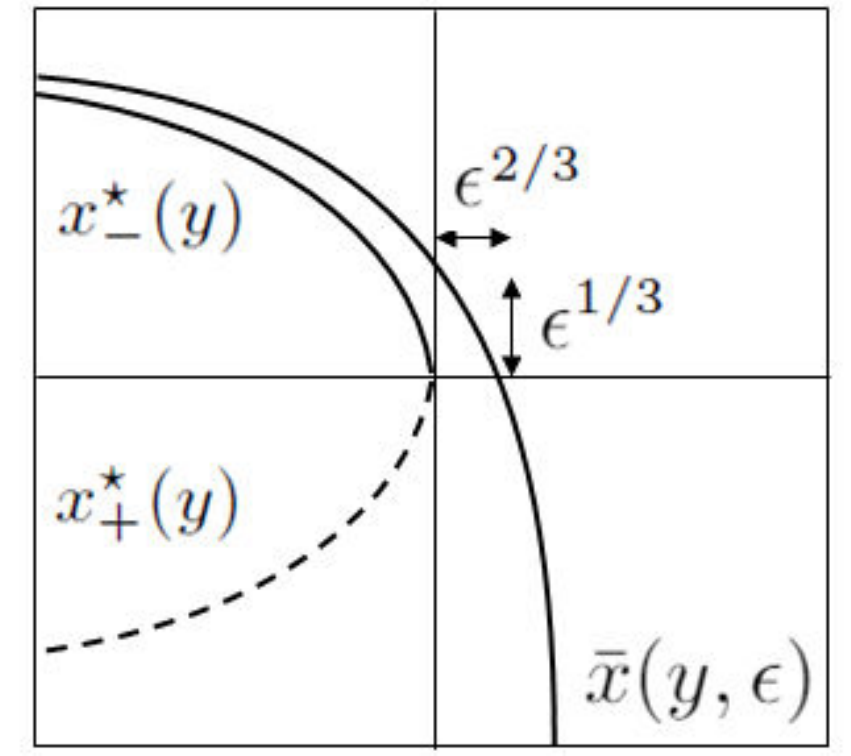}
    \caption{Solutions track the stable branch at a distance $O(\epsilon ^{1/3})$ \cite{Gentz06}.}
    \label{onedimensional}
\end{figure}

It has been shown in Theorem 3.10 \cite{Laing10} and illustrated in Fig. \ref{onedimensional} that near the SNB, the invariant manifold $\bar{x}(y,\epsilon)$ of (\ref{normal form near SNB}) will first cross the axis $y=0$ for a $x$ of $O(\epsilon^{1/3})$, then the axis $x=0$ for a $y$ of $O(\epsilon^{2/3})$. 


If we incorporate the stochasticity and consider (\ref{general slow fast SDE}), the impacts of the stochastic perturbation to the slow-fast system near the SNB are summarized in the following theorem.

\textit{Theorem 2 \cite{Gentz06}, \cite{Laing10}:} Consider the slow-fast system (\ref{general slow fast SDE}) in which $n_x=1$, $n_y=1$, assume that the conditions \textit{A1-A4} are satisfied, then the following hold:
\begin{enumerate}
    \item If $\sigma<\sigma_c=\sqrt{\epsilon}$, the sample paths remain in a neighborhood ${\cal B}(h)$ of the deterministic solution with probability larger than $1-O(e^{-h^2/{2\sigma^2}})$ for all $h$ up to order $\sqrt{\epsilon}$, as long as $y(t)<c_1\epsilon^{2/3}$.  
    \item If $\sigma>\sigma_c=\sqrt{\epsilon}$, then sample paths are likely to cross the unstable branch and reach negative values of order 1 for $y(t)$ of order $-\sigma^{4/3}$. \textcolor{BrickRed}{The probability that the early transition does not happen is of order $e^{-O(\sigma^2/{\epsilon |log{\sigma}|})}$.}
\end{enumerate}

\begin{figure}[!b]
\centering
\subfloat[$\sigma<\sqrt{\epsilon}$]{\includegraphics[width=1.65in]{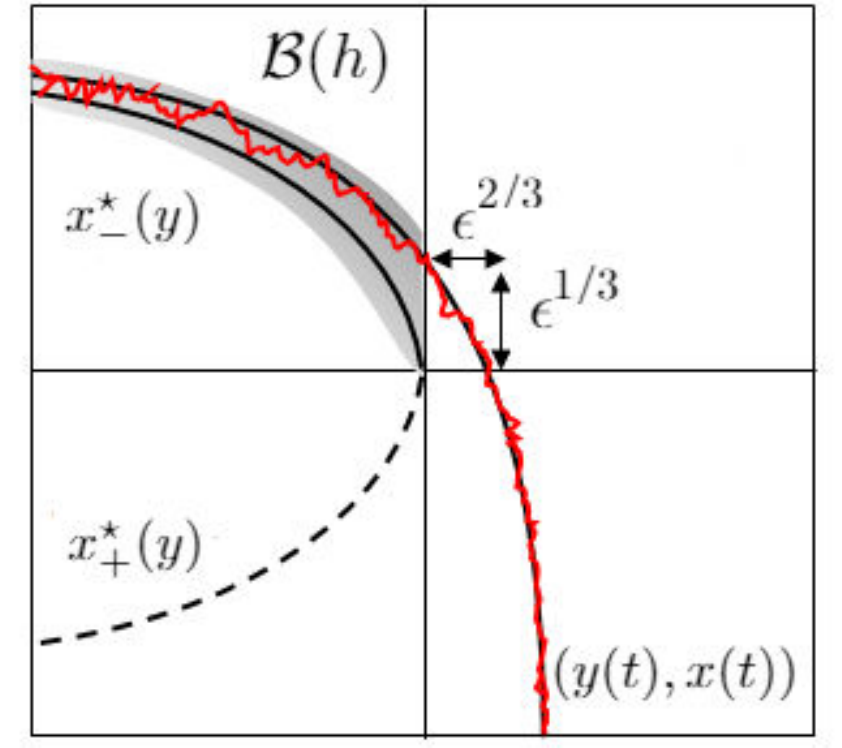}
\label{weak}}
\hfil
\subfloat[$\sigma>\sqrt{\epsilon}$]{\includegraphics[width=1.65in]{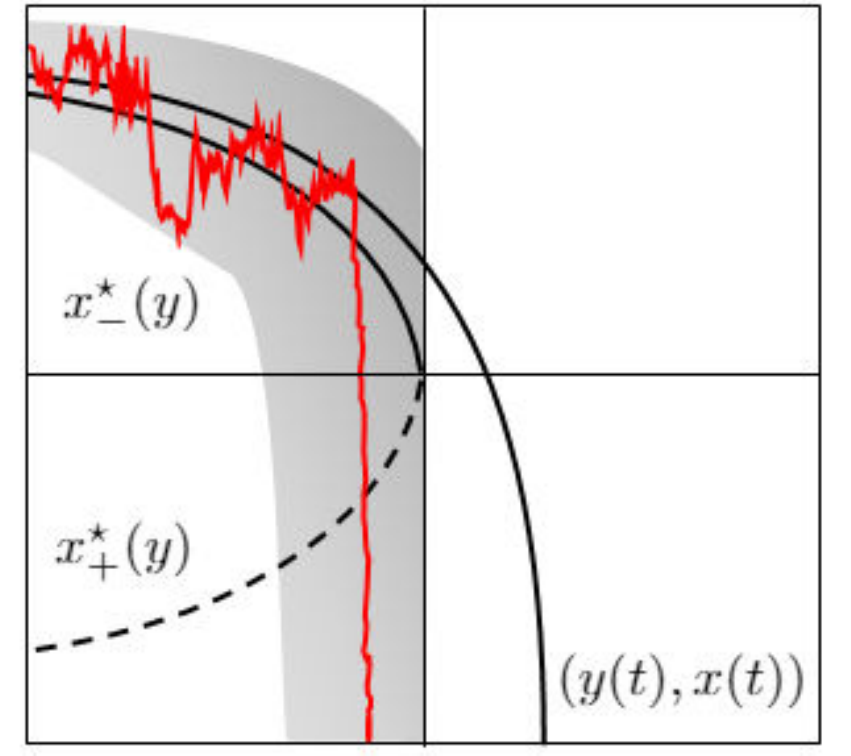}
\label{strong}}
\caption{An illustration for Theorem 2.} 
\label{weak_strong_regimes}
\end{figure}
\textit{Theorem 2} implies that if $\sigma<\sqrt{\epsilon}$, the situation is similar to the deterministic system: sample paths feel the bifurcation after a delay of $O(\epsilon^{2/3})$. 
However, if $\sigma>\sqrt{\epsilon}$, 
sample paths feel the bifurcation some time before it happens and jump to negative $x$. An illustration is given in Fig. \ref{weak_strong_regimes}. This result indicates that the stochastic perturbation may greatly affect the sample path qualitatively near the SNB, and thus needs to be carefully considered. Particularly, the influence of the stochastic perturbation depends on the relative size of the parameters $\sigma$ and $\epsilon$.


\section{Analytical Study of the Impacts of Stochastic Load Variation}
\label{3}
We consider the stochastic power system dynamic model given in (\ref{eq:sde_model_u}) with slowly varying parameters $\bm{p}$:
\begin{equation}
\label{eq:sde_model_u_1}
\dot{\bm{u}} = \bm{G}(\bm{u},\bm{p},\epsilon t)+\sigma B \bm{\xi} \\
\end{equation}
where $\bm{p}$ are the  real power $p$ and reactive power $q$ of loads, which
experience a gradual increase with respect to time:
\begin{equation}
\label{eq:PQincrease}
\begin{gathered}
p=p_0(1+\lambda(\epsilon t)) \\
q=q_0(1+\lambda(\epsilon t))
\end{gathered}
\end{equation}
where $\lambda$ is a slowly increasing loading factor. \textcolor{BrickRed}{The maximum value $\lambda$ before the SNB occurs multiplied by $p_0$  corresponds to the voltage stability margin $S$, i.e. $S=\lambda p_{0}$ at the SNB.} 

Without stochastic load variations, (\ref{eq:sde_model_u_1}) becomes a deterministic system:
\begin{equation}
\label{eq:ode_model_u_1}
  \dot{\bm{u}} = \bm{G}(\bm{u},\bm{p},\epsilon t)
\end{equation}

We will discuss next how, in comparison to the deterministic system (\ref{eq:ode_model_u_1}), the stochastic term $\sigma B \bm{\xi}$ will affect the trajectory of (\ref{eq:sde_model_u_1}) near the normal operation point and near the SNB point, respectively.

\subsection{Concentration of Sample Paths around the Stable Component of the Constraint Manifold}\label{sectionconcentrationofpaths}
To present the concentration results, we define a manifold:
\begin{equation}
\label{eq:N0}
{\cal N}_0=\{\bm{u^{\star}}(\bm{p},\epsilon t):\bm{G}(\bm{u^{\star}}(\bm{p},\epsilon t),\bm{p},\epsilon t)=\bm{0}\}
\end{equation}
and an invariant manifold $\epsilon$-away from ${\cal N}_0$:
\begin{equation}
\label{eq:Ne}
{\cal N}_\epsilon=\{\bar{\bm{u}}(\bm{p},\epsilon t)=\bm{u^{\star}}(\bm{p},\epsilon t)+O(\epsilon)\}
\end{equation}
In addition, we define an ellipsoidal layer ${\cal N}(h)$ surrounding the invariant manifold ${\cal N}_{\epsilon}$:
\begin{equation}
{\cal N}(h) = \{( \bm{u},t):\langle \bm{u}-\bm{\bar{u}}(\bm{p},\epsilon t),\bm{\bar{U}}(\bm{p},\epsilon t)^{-1}( \bm{u-\bar{u}}(\bm{p}, \epsilon t)) \rangle<h^2 \}
\label{neighborhood_u}
\end{equation}
where $\bm{\bar{U}}(\bm{p},\epsilon t)$ is defined in Appendix \ref{appendix_crosssection}.
Then we have the following theorem showing that under some mild conditions, the sample paths of the stochastic system (\ref{eq:sde_model_u_1}) are concentrated in the ellipsoidal layer ${\cal N}(h)$. 


\textit{Theorem 3:} Consider the system (\ref{eq:sde_model_u_1}), for some fixed $\epsilon_0>0$, $h_0>0$, $\delta_0>0$, a time $t_1$ of order $|\log h|$ such that if the following two conditions are satisfied:
\begin{enumerate}
    \item The slow manifold ${\cal N}_0$ is a stable component of the constraint manifold;
    \item The initial condition $\bm{u}(0)\in {\cal N}(\delta_0)$,
\end{enumerate}
then for $t\geq t_1$, the sample path of the stochastic system (\ref{eq:sde_model_u_1}) satisfies the following relation:
\begin{eqnarray}
   && \mathbb{P}\{\exists t_2\in[t_1,t]: \textcolor{BrickRed}{(\bm{u},t)}\not\in{\cal N}(h)\}\nonumber\\
   &&\leq C_{n_{\bm{u}}}(t,\epsilon) e^{\frac{-h^2}{2\sigma^2}(1-O(h)-O(\epsilon))}\label{eq:probabilityofleaving}
\end{eqnarray}
for all $\epsilon\leq \epsilon_0$, $h\leq h_0$, where the coefficient  $C_{n_{\bm{u}}}(t,\epsilon)$ is linear in time.

Proof: The stochastic dynamic power system model (\ref{eq:sde_model_u_1}) can be represented as a stochastic slow-fast system:
\begin{eqnarray}
\label{eq:sde_model_u_slow fast}
\dot{\bm{u}}& =& \bm{G}(\bm{u},\bm{p},y)+\sigma B \bm{\xi} \nonumber\\
\dot{y}&=&\epsilon
\end{eqnarray}
where $\textcolor{BrickRed}{y(t)=\epsilon t}$. If passed to the slow time scale $s=\epsilon t$, it takes the form:
\begin{eqnarray}
\label{eq:u_slow fast_slowtimescale}
{\bm{u}^\prime}& =& \frac{1}{\epsilon}\bm{G}(\bm{u},\bm{p},y)+\frac{\sigma}{\sqrt{\epsilon}} B \bm{\xi} \nonumber\\
{y^\prime}&=&1
\end{eqnarray}
where $^\prime=\frac{d}{ds}$, $\textcolor{BrickRed}{y(s)=s}$. Its deterministic counterpart is:
\begin{eqnarray}
\label{eq:u_det_slow fast_slowtimescale}
{\bm{u}^\prime}& =& \frac{1}{\epsilon}\bm{G}(\bm{u},\bm{p},y) \nonumber\\
{y^\prime}&=&1
\end{eqnarray}
The slow manifold of (\ref{eq:u_det_slow fast_slowtimescale}) can be represented as  $\{\bm{u^{\star}}(\bm{p},y):\bm{G}(\bm{u^{\star}}(\bm{p},y),\bm{p}, y)=\bm{0}\}$, which is exactly the same as ${\cal N}_0$ defined in (\ref{eq:N0}). Likewise, the $\epsilon$-invariant manifold of (\ref{eq:u_det_slow fast_slowtimescale}) that attracts the nearby solutions exponentially fast is ${\cal N}_\epsilon$ as defined in (\ref{eq:Ne}). As a result, by \textit{Theorem 4} in Appendix \ref{appendix_samplepath}, the relationship (\ref{eq:probabilityofleaving}) is satisfied. This completes the proof.

\textit{Theorem 3} implies that if $h\gg\sigma$, i.e., the depth of the ellipsoidal layer is much greater than the noise intensity $\sigma$, the right hand side of (\ref{eq:probabilityofleaving}) becomes very small, i.e., the sample paths of (\ref{eq:sde_model_u_1}) will unlikely leave ${\cal N}(h)$.
Note that the detailed expression of $C_{n_{\bm{u}}}(t,\epsilon)$ is not of interest (see Theorem 2.4 in \cite{Gentz03}). The key point is that $C_{n_{\bm{u}}}(t,\epsilon)$ is independent of $h$ and $\sigma$. As a result, the probability of the sample path to leave layer ${\cal N}(h)$ decays exponentially as $h$ increases.

The above results reveal that the trajectory of the stochastic power system model (\ref{eq:sde_model_u_1}) will be likely concentrated in a small neighborhood of the trajectory of the deterministic power system model (\ref{eq:ode_model_u_1}) when the power system is operating in normal conditions. 
A natural question to ask is what will happen if the system is getting close to the SNB point. 



\subsection{The Impact of Noise on Saddle-Node Bifurcation}
\label{dynamic_snb}
We have shown that the stochastic power system model (\ref{eq:sde_model_u_1}) can be represented as a stochastic slow-fast system (\ref{eq:u_slow fast_slowtimescale}). By \textit{Theorem 3}, it is seen that the shape of the concentration neighborhood $\cal{N}$$(h)$ depends on $\epsilon$ (see the detailed expression of $\bm{\bar{U}}(\bm{p},\epsilon t)$ in Appendix \ref{appendix_crosssection}), whereas the depth of ${\cal N}(h)$ is related to $\sigma$.

Mathematically, $\epsilon$ describes the decoupling between the slow and the fast dynamics, while $\sigma$ describes the intensity of the stochastic perturbations.
In the context of power system voltage stability study, $\sigma$ describes the intensity of load fluctuations and $\epsilon$ describes the \textcolor{BrickRed}{load power variation speed}. 

\begin{figure}[!tb]
\vspace{-5pt}
\centering
\subfloat[Weak regime $\sigma<\sqrt{\epsilon}$]{\includegraphics[width=1.4in]{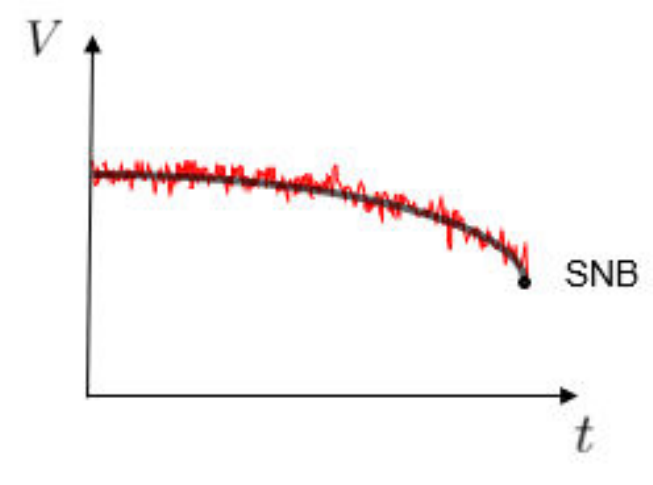}
\label{weakregime}}
\hfil
\subfloat[Strong regime $\sigma>\sqrt{\epsilon}$]{\includegraphics[width=1.4in]{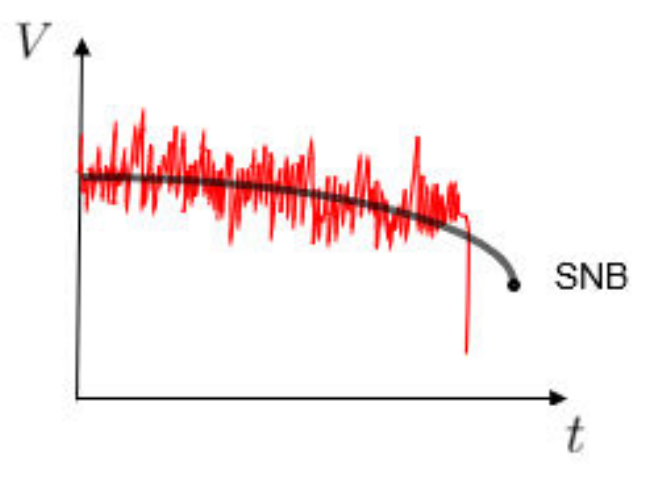}
\label{strongregime}}
\caption{An illustration of the weak and strong noise regimes on the time evolution of voltage.}
\label{regimes}
\end{figure}

It is revealed in \textit{Theorem 2} that the randomness may greatly affect the dynamics of a system near SNB. Particularly, a trade-off relationship between $\sigma$ and $\epsilon$ exists, based on which a weak noise regime and a strong noise regime can be defined and are illustrated in Fig. \ref{regimes}:  
\begin{enumerate}
\item \textit{Weak Noise Regime:} if  $\sigma<\sqrt{\epsilon}$, 
the trajectories of the stochastic power system model (\ref{eq:sde_model_u_1}) will continue  remaining concentrated in the neighborhood ${\cal N}(h)$ 
surrounding the deterministic solution (\ref{eq:ode_model_u_1}).
\item \textit{Strong Noise Regime:} if $\sigma>\sqrt{\epsilon}$, 
it becomes likely that the samples paths may reach the SNB point earlier.
\end{enumerate}

The above results indicate that the distinction between the weak noise regime and the strong noise regime depends not only on the stochastic perturbation intensity $\sigma$, but also on the \textcolor{BrickRed}{load power variation speed} $\epsilon$. Particularly, the influence of the \textcolor{BrickRed}{load power variation speed} on the dynamic voltage stability margin has not been reported in previous work. In addition, the trade-off relationship between the two parameters $\sigma$ and $\epsilon$ implies a potential of using \textcolor{BrickRed}{energy storage systems} to maintain the same level of the voltage stability margin under high uncertainty.

\textcolor{BrickRed}{Furthermore, a metric may be derived from the theoretical result to estimate the reduction of voltage stability margin due to the randomness. 
Specifically, by \textit{Theorem 2}, in the strong noise regime, the sample paths are likely to reach the bifurcation earlier and cross the unstable branch for time of order $-\sigma^{4/3}$. That being said, assuming that the intensity of the stochastic load fluctuations $\sigma$ and the deterministic load margin $S_{det}$ are known, \textit{Theorem 2} implies that the stochastic power system model trajectories may feel the bifurcation and collapse earlier, leading to a decrease of order $\sigma^{4/3} \times S_{det}$ (MW) in terms of the voltage stability margin. Further study to  define a more concrete metric is needed.}

\section{Numerical Results}\label{4}

In this section, we will numerically study the impacts of randomness on the voltage stability margin to illustrate and validate the analytical results in Section \ref{3}. We will firstly show that there exists a weak/strong noise regime in which the stochastic perturbation will not/will greatly affect the dynamic voltage stability margin. Secondly, we will investigate the impacts of the \textcolor{BrickRed}{load power variation speed} $\epsilon$ on the dynamic voltage stability margin as suggested by \textit{Theorem 2}, yet has not been studied in previous work. Lastly, we will explore the trade-off relationship between the stochastic perturbation intensity $\sigma$ and the \textcolor{BrickRed}{load power variation speed} $\epsilon$ and discuss its significant insights in control design.

Numerical study and Monte Carlo time domain simulations 
were carried out
on the IEEE 14-bus system in which an aggregated dynamic load (\ref{eq:erl_power})-(\ref{eq:exp_rec_load_sde}) was added to Bus 9.
 Ornstein-Uhlenbeck load fluctuations with $\alpha_{i}=1,  \beta_{i}=\sqrt{2\alpha_{i}}=\sqrt{2}$
were applied to the aggregated load model as in (\ref{eq:ouprocess}).
PSAT Toolbox \cite{Milano} was used to perform all the simulations. Euler-Maruyama method was used to generate the Ornstein-Uhlenbeck process and the integration step size was $\Delta t=0.05$s. Active and reactive power limits of generators have been considered. \textcolor{BrickRed}{The singularity of the state matrix based on the reciprocal of its condition number (Matlab command: rcond) was used as a criterion to identify the SNB point and calculate the dynamic load margin. The value 0.1 has been used as a threshold for the reciprocal of the state matrix's condition number to detect its singularity.}

\subsection{The Existence of Weak and Strong Noise Regimes}
\label{resultsA}

By the analytical results in Section \ref{dynamic_snb}, given fixed $\epsilon$, i.e., a constant \textcolor{BrickRed}{load power variation speed}, a fluctuation intensity $\sigma$ may lead to a smaller voltage stability margin if $\sigma>\sqrt{\epsilon}$. To show this, 
we apply various intensities $\sigma_1=0.05, \sigma_2=0.10, \sigma_3=0.15$ for fixed $\epsilon$, i.e., the load at Bus 4 increases by 2\% of its nominal power every 0.4s, \textcolor{BrickRed}{corresponding to a load power variation speed of $2$ MW/s}. 
The mean 
and variance 
of the stability margin \textcolor{BrickRed}{$S$} 
for each case estimated from 1000 Monte-Carlo simulations  are presented in Table \ref{tb:case1}.
Note that the margin for the deterministic case where $\sigma=0$ is \color{BrickRed}${S_{det}}=542.75$ MW\color{black}. The percentage difference of the mean load margin with respect to \textcolor{BrickRed}{$S_{det}$} is also given in the last row of Table \ref{tb:case1}.

\begin{figure}[!b]
\centering
\includegraphics[width=1.9in,keepaspectratio=true,angle=0]{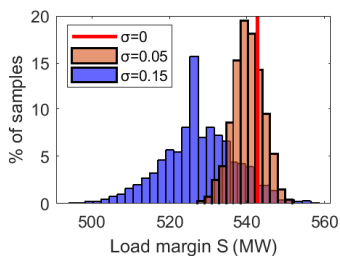}
\caption{\color{black}The distribution of the margin for various fluctuation intensities when \color{BrickRed}{the load power variation speed is 2 MW/s.}\color{black}}
\label{histo_allsigma_0.4s_1}
\end{figure}

As expected, the mean value of the load margin decreases as the intensity of load fluctuation increases.
Particularly, Fig. \ref{histo_allsigma_0.4s_1} depicts the distributions of the margin when different fluctuation intensities are applied. It is obvious that the histograms are shifted to the left when it comes to larger $\sigma$. When $\sigma=0.05$, 
the impact of noise on the margin size is almost negligible. In contrast, when $\sigma=0.15$, the shrinking of the voltage stability margin is pronounced. 
If, for instance, $2\%$ margin reduction with respect to \textcolor{BrickRed}{$S_{det}$} is the threshold determining if a decrease of the stability margin occurs, 
i.e., the system is in strong noise regime or not, we see that when $\sigma=0.05$ or $0.10$, the system is in the weak noise regime; if $\sigma=0.15$, the system is in the strong noise regime.

\begin{table}[!b]
\centering
  \caption{The statistics of the voltage stability margin \textcolor{BrickRed}{$S$} for various fluctuation intensities when \color{BrickRed}{load speed is 2MW/s}\color{black}
  }\label{tb:case1}
  \begin{tabular}{|c|c|c|c|}
\hhline{|-|-|-|-|}
{Fluctuation Intensity $\sigma$}&0.05&0.10&0.15\\
 \hline
\textcolor{BrickRed}{$\mu= $ Mean $S$ (MW)}\textcolor{black}& 539.78&534.24&527.67\\
\hline
\textcolor{BrickRed}{Variance $S$}\color{black}
& 16.17&51.09&94.13\\
\hline
\color{BrickRed}{$\frac{\mu-S_{det}}{S_{det}}$}\color{black}
& -0.55\%&-1.57\%&-2.78\%\\
\hhline{|-|-|-|-|}
  \end{tabular}
\end{table}

\subsection{The Impacts of the \textcolor{BrickRed}{Load Power Variation Speed}\color{black}}\label{resultsB} 

\begin{figure}[!t]
\centering
\includegraphics[width=1.85in,keepaspectratio=true,angle=0]{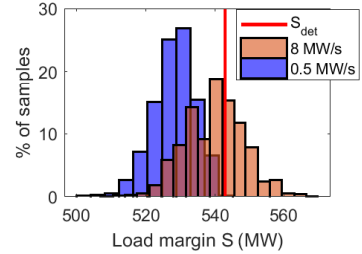}
\caption{The distribution of the load margin when \textcolor{BrickRed}{the load power variation speed is $8$ MW/s and $0.5$ MW/s }
with fixed fluctuation intensity $\sigma=0.10$.
A slower \textcolor{BrickRed}{load power variation speed} leads to a smaller load margin.}
\label{histo_speeds_sigma0.10}
\end{figure}

We have observed that 
different intensities of load fluctuation will make the system lie in different noise regimes, indicating the impacts of randomness on the size of the dynamic load margin. However, 
the separation of different noise regimes shown in Section \ref{dynamic_snb} suggests that $\epsilon$, the \textcolor{BrickRed}{load power variation speed}, also plays an important role. To investigate the impacts of the \textcolor{BrickRed}{load power variation speed}, we increase the power of the load at Bus 4 at different speeds as in (\ref{eq:PQincrease}), namely 2\% of $p_{0}$ and $q_{0}$ every 0.1s, 0.4s, 0.9s and 1.6s, \textcolor{BrickRed}{corresponding to the load power variation speeds $8$ MW/s, $2$ MW/s, $0.9$ MW/s and $0.5$ MW/s} respectively, while keeping $\sigma=0.10$. Note that the time interval between each power change is $O(\frac{1}{\epsilon})$. The statistics of the voltage stability margin are shown in Table \ref{tb:case2}. Moreover, Fig. \ref{histo_speeds_sigma0.10} presents the distributions of the margin size for different \textcolor{BrickRed}{load power variation speeds} when the fluctuation intensity of the aggregated load is $\sigma=0.10$. 

If we still regard $2\%$ reduction of the voltage stability margin \textcolor{BrickRed}{comparing to the deterministic case $S_{det}$} to be the threshold, we see that if the load increases at a fast speed, say \textcolor{BrickRed}{$8$ MW/s or $2$ MW/s}, the system lies in the weak noise regime, indicating that the stochastic load fluctuation does not greatly affect the margin size; however, if the load increases at a slow speed, say \textcolor{BrickRed}{$0.9$ MW/s or $0.5$ MW/s}, the system is in the strong noise regime, where a more than $2\%$ reduction of the margin size is expected.

The above results can be well explained by the analytical results presented in Section \ref{dynamic_snb}. For fixed $\sigma$, a smaller  $\sqrt{\epsilon}$, i.e., a slower \textcolor{BrickRed}{load power variation speed}, results in a smaller threshold value separating the two regimes, which therefore leads to a smaller weak noise regime and a larger strong noise regime. 
That being said, for fixed load fluctuation intensity, a slower \textcolor{BrickRed}{load power variation speed} may lead to an earlier voltage collapse and a greater reduction of the voltage stability margin size.

These interesting results about the influence of the time evolution property of driving parameters on the dynamic voltage stability margin have not been discussed in previous voltage stability study. 
However, they demonstrate the importance of incorporating the time evolution property in assessing the voltage stability margin especially under a high uncertainty level.
In addition, significant insights can be extracted from the trade-off relationship between $\epsilon$ and $\sigma$ as discussed next. 

\subsection{Important Insights behind $\sigma< \sqrt{\epsilon}$:}
From the previous analysis, we have seen that both the fluctuation intensity $\sigma$ and the \textcolor{BrickRed}{load power variation speed} $\epsilon$ may affect the size of the voltage stability margin. Inspired by this observation, we have estimated the statistics of the margin size for different combinations of fluctuation intensity $\sigma$ and load power variation speed $\epsilon$, as shown in Tables \ref{meanvalues}-\ref{decrease}. Particularly, we use $^\star$ to denote the results from Section \ref{resultsA}, and $^{\diamond}$ to denote the results from Section \ref{resultsB}.

\begin{figure}[!htb]
\centering
\subfloat[$\sigma=0.05$]{\includegraphics[width=1.68in]{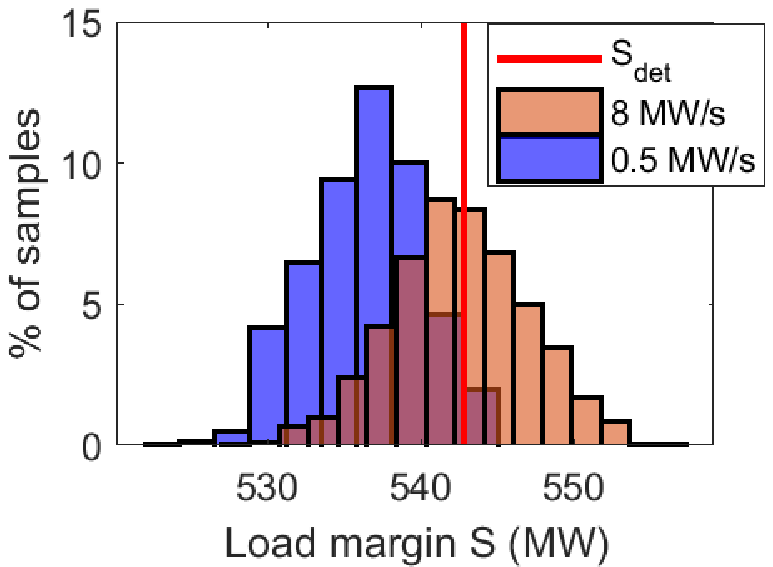}
\label{histo_speeds_sigma0.05}}
\hfil
\subfloat[$\sigma=0.15$]{\includegraphics[width=1.68in]{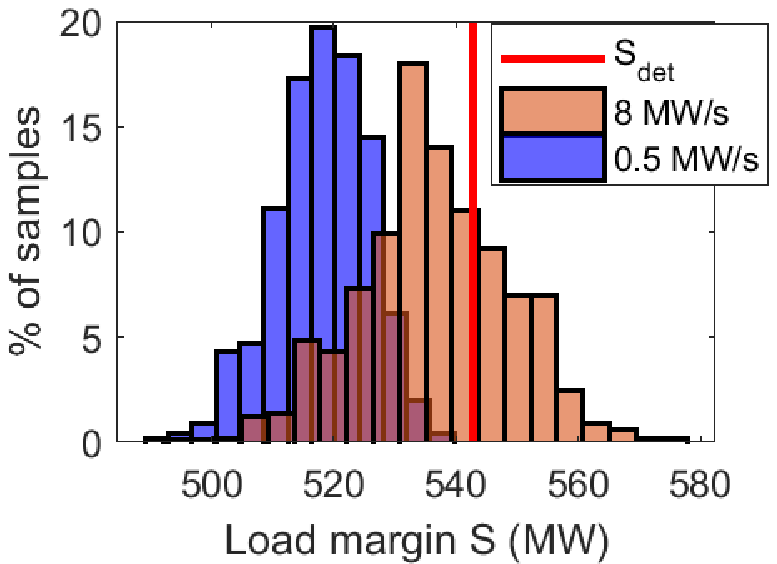}
\label{histo_speeds_sigma0.15}}
\caption{(a) The distribution of the load margin when \color{BrickRed}{the load power variation speed is 8 MW/s or 0.5 MW/s }\color{black}
with a fluctuation intensity of 0.05;
(b) The distribution of the load margin when \color{BrickRed}{the load power variation speed is 8 MW/s or 0.5 MW/s }\color{black}
with a fluctuation intensity of 0.15. A slower \textcolor{BrickRed}{load power variation speed} or a larger fluctuation intensity leads to a smaller load margin.} 
\label{histo_speeds}
\end{figure}

\begin{figure}[!htb]
\centering
\includegraphics[width=3in,keepaspectratio=true,angle=0]{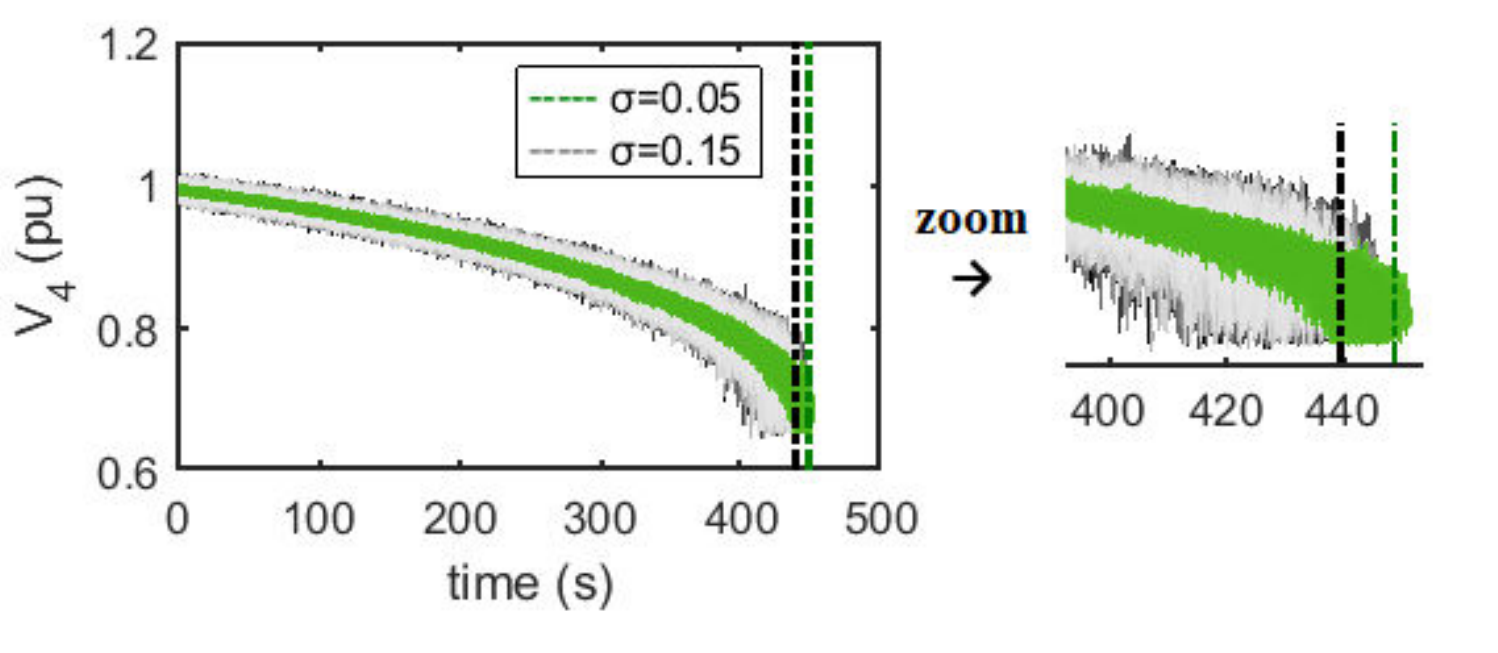}
\caption{$|V_4|$ from 1000 trajectories of different fluctuation intensities $\sigma$ when \color{BrickRed}{the load power variation speed is 0.5 MW/s. }\color{black}
The straight lines denote where 96\% of the trajectories have collapsed.}\label{1000traj_1.6s_96}
\end{figure}

Fig. \ref{histo_speeds} illustrates the distributions of the dynamic load margin for different fluctuation intensities when different speeds are applied. As expected, we observe a shift to the left when it comes to larger fluctuation intensities or slower \textcolor{BrickRed}{load power variation speeds}. Fig. \ref{1000traj_1.6s_96} presents the voltage magnitude at Bus 4 for different intensities yet for the same speed. 
The sample paths are concentrated around the $\sigma$-neighbourhood of the deterministic solution.  Thus, Fig. \ref{1000traj_1.6s_96} corroborates the analytical results shown in Section \ref{sectionconcentrationofpaths} that the depth of the concentration neighborhood depends on $\sigma$. 
Besides, it can be seen that for large $\sigma$, the majority of the trajectories will collapse earlier than those with smaller $\sigma$.

\begin{table}[!t]
\centering
  \caption{The statistics of the voltage stability margin \textcolor{BrickRed}{$S$} for various \textcolor{BrickRed}{load power variation speeds} with fixed fluctuation intensity $\sigma=0.10$}\label{tb:case2}
  \vspace{-3pt}
  \begin{tabular}{|c|c|c|c|c|}
\hhline{|-|-|-|-|-|}
Load speed \textcolor{BrickRed}{(MW/s)} & \textcolor{BrickRed}{8}&\textcolor{BrickRed}{2} & \textcolor{BrickRed}{0.9} & \textcolor{BrickRed}{0.5}\\
 \hline
\textcolor{BrickRed}{$\mu= $ Mean $S$ (MW)}& 540.00&534.24&530.67&527.76\\
\hline
\textcolor{BrickRed}{Variance $S$} & 74.30&51.09&37.08&33.32\\
\hline
\color{BrickRed}{$\frac{\mu-S_{det}}{S_{det}}$} & -0.51\%&-1.57\%&-2.23\%&-2.76\%\\
\hhline{|-|-|-|-|-|}
  \end{tabular}
\end{table}

\begin{table}[!b]
\centering
  \caption{The mean value of the voltage stability margin
  }\label{meanvalues}
  \begin{tabular}{|c|c|c|c|c|}
\hhline{|-|-|-|-|-|}
\diagbox[width=1.34in, height=0.34in]{Fluc. Intensity $\sigma$}{Load speed \textcolor{BrickRed}{(MW/s)}}&\textcolor{BrickRed}{8}&\textcolor{BrickRed}{2}&\textcolor{BrickRed}{0.9}&\textcolor{BrickRed}{0.5}\\
  \hline
   0.05&542.16&\color{amethyst}539.78$^\star$\color{black}&538.00&\color{blue}536.38\color{black} \\
   \hline
   0.10&\color{amethyst}540.00$^{\diamond}$\color{black}&\color{blue}534.24$^{\diamond\star}$\color{black}&530.67$^{\diamond}$&527.76$^{\diamond}$\\
    \hline
   0.15&536.57&527.67$^\star$&522.29&518.41 \\
\hhline{|-|-|-|-|-|}
  \end{tabular}
  \vspace{-7pt}
\end{table}

\begin{table}[!htb]
\centering
  \caption{The variance of the voltage stability margin} \label{variances}
  \begin{tabular}{|c|c|c|c|c|}
\hhline{|-|-|-|-|-|}
\diagbox[width=1.32in, height=0.32in]{Fluc. Intensity $\sigma$}{Load speed \textcolor{BrickRed}{(MW/s)}}&\textcolor{BrickRed}{8}&\textcolor{BrickRed}{2}&\textcolor{BrickRed}{0.9}&\textcolor{BrickRed}{0.5}\\
  \hline
   0.05&20.54&16.17$^\star$&12.19&11.15 \\
   \hline
   0.10&74.30$^{\diamond}$&51.09$^{\diamond\star}$&37.08$^{\diamond}$&33.32$^{\diamond}$\\
    \hline
   0.15&151.91&94.13$^{\star}$&73.83&57.83 \\
\hhline{|-|-|-|-|-|}
  \end{tabular}
  \vspace{-7pt}
\end{table}

\begin{table}[!htb]
\centering
  \caption{The percentage difference of the mean load margin comparing to the deterministic case \textcolor{BrickRed}{($\frac{\mu-S_{det}}{S_{det}}$)}
  }\label{decrease}
  \begin{tabular}{|c|c|c|c|c|}
\hhline{|-|-|-|-|-|}
\diagbox[width=1.3in, height=0.32in]{Fluc. Intensity $\sigma$}{Load speed \textcolor{BrickRed}{(MW/s)}}&\textcolor{BrickRed}{8}&\textcolor{BrickRed}{2}&\textcolor{BrickRed}{0.9}&\textcolor{BrickRed}{0.5}\\
  \hline
   0.05&-0.11\%&-0.55\%$^\star$&-0.88\%&-1.17\% \\
   \hline
   0.10&-0.51\%$^{\diamond}$&-1.57\%$^{\diamond\star}$&-2.23\%$^{\diamond}$&-2.76\%$^{\diamond}$\\
    \hline
   0.15&-1.14\%&-2.78\%$^{\star}$&-3.77\%&-4.48\%\\
\hhline{|-|-|-|-|-|}
  \end{tabular}
\end{table}


\begin{figure}[!t]
\centering
\subfloat[Base case (yellow): the fluctuation intensity $\sigma=0.10$, \color{BrickRed}{the load power variation speed is 8 MW/s.}\color{black}]
{\includegraphics[width=1.65in]{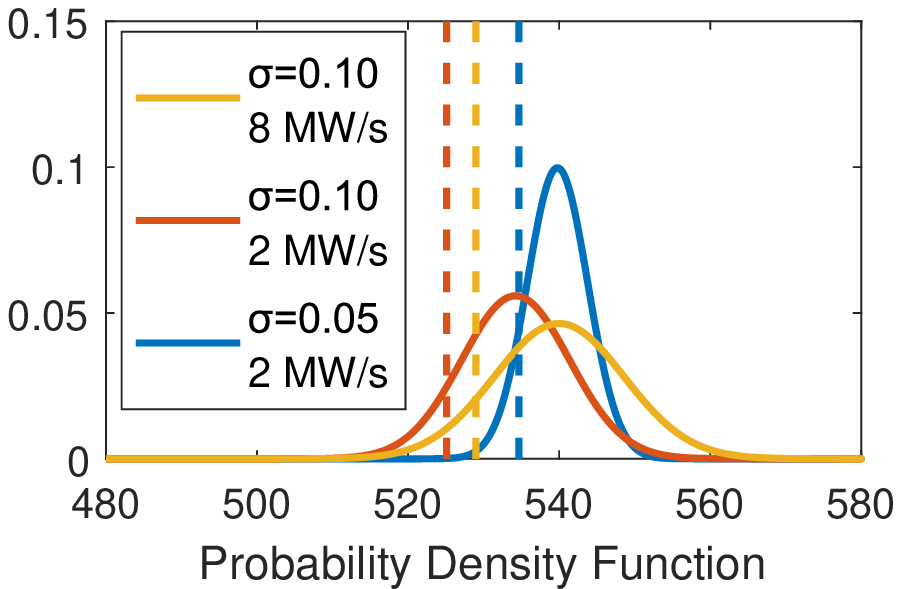}
\label{varianceincrease}}
\hfil
\subfloat[Base case (yellow): the fluctuation intensity $\sigma=0.05$, \color{BrickRed}{the load power variation speed is 0.5 MW/s}\color{black}.]
{\includegraphics[width=1.65in]{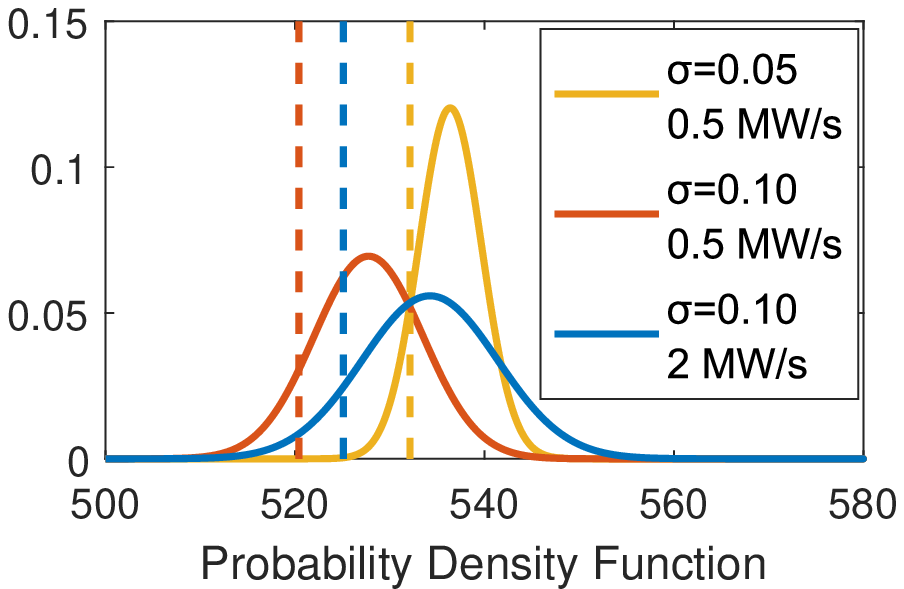}
\label{speedincrease}}
\caption{Illustration of using \textcolor{BrickRed}{energy storage systems} to keep the same level of voltage stability margin. The straight lines indicate the lower bound of 90\% confidence interval.}

\end{figure}

Another interesting observation from Tables \ref{meanvalues}-\ref{decrease} is that if the \textcolor{BrickRed}{load power variation speed} is slower than expected while the same level of stability margin is desired, energy storage may be applied to smooth out the variance of the load. 
Such results can be found in the entries of Table \ref{meanvalues} denoted in purple and in blue, respectively. For instance, with $\sigma=0.10$ and a \textcolor{BrickRed}{load speed of 8 MW/s, i.e., the load increases per 2\% of its base power every $0.1$s}, we have a margin size of $540.00$ MW (the PDF with 90\% confidence are denoted in yellow in Fig. \ref{varianceincrease}). However, if the \textcolor{BrickRed}{load power actually increases every $0.4$ s, corresponding to the slower load power variation speed 2 MW/s}, the margin reduces to $534.24$ MW (the PDF with 90\% confidence are denoted in red in Fig. \ref{varianceincrease}). To maintain the same level of margin size as of \color{BrickRed}{the speed $8$ MW/s}\color{black}, we need to adopt control measures to reduce the load fluctuation intensity to $\sigma=0.05$, which will result in a margin of $539.78$ MW (the PDF with 90\% confidence are denoted in blue in Fig. \ref{varianceincrease}). Although it is intuitive that reducing the variance of random inputs may help the voltage stability, the relation $\sigma < \sqrt{\epsilon}$ provides important guidelines regarding how much reduction in terms of the fluctuation intensity is needed to maintain the same level of margin size.
In this case, when $\epsilon$, i.e., the threshold separating the weak and noise regimes, decreases by 4 times, $\sigma$ needs to be reduced by $\sqrt{4}=2$ times.

Likewise, controlling the \textcolor{BrickRed}{load power variation speed} using \textcolor{BrickRed}{energy storage systems} may also compensate for unexpected increasing fluctuation intensity. For example,
with fluctuation intensity $\sigma=0.05$ and \textcolor{BrickRed}{a load that increases every $1.6$s, corresponding to the speed $0.5$ MW/s}, we have a margin size of $536.38$ MW (the PDF with 90\% confidence are denoted in yellow in Fig. \ref{speedincrease}). However, if the actual fluctuation intensity is $\sigma=0.10$, the margin reduces to $527.76$ MW (the PDF with 90\% confidence are denoted in red in Fig. \ref{speedincrease}). To maintain the same level of margin size, we can use \textcolor{BrickRed}{energy storage systems} to increase the load power variation speed to \textcolor{BrickRed}{$2$ MW/s, i.e. the load increases every $0.4$s} (e.g., by increasing the charging power of \textcolor{BrickRed}{energy storage systems} \cite{Su13}), which will result in a margin of $534.24$ MW (the PDF with 90\% confidence are denoted in blue in Fig. \ref{speedincrease}). 

\section{Conclusions and Perspectives}\label{5}

This paper has presented an analytical investigation of the impacts of stochastic load fluctuations on the size of the dynamic voltage stability margin using bifurcation theory. Through analytical study and systematic numerical study, it has been shown that two parameters, 
namely the load fluctuation intensity and the load power variation speed, will affect the size of the voltage stability margin. \textcolor{BrickRed}{Particularly, the influence of the time evolution property of the driving parameters (e.g., the load power variation speed) on power system voltage stability in the presence of uncertainty is revealed}. It has been observed that a slower \textcolor{BrickRed}{load power variation speed} or a larger load variation intensity may lead to a smaller voltage stability margin. Therefore, it is crucial to consider 
both factors to accurately assess the voltage stability. Additionally, by exploiting bifurcation theory of the stochastic dynamic system, the trade-off relationship between the two parameters has been revealed and validated. Simulation results provide significant insights regarding the potential of using \textcolor{BrickRed}{energy storage systems} to maintain the voltage stability of the grid under high uncertainty level.
It is worth mentioning that such outcomes cannot be observed by using static or deterministic approaches, which in turn reinforces the importance of carrying out dynamic and stochastic approaches in voltage stability analysis, especially considering the increasing degree of uncertainty in modern power systems due to the integration of \textcolor{BrickRed}{renewable energy sources}. Future work will focus on implementing an \textcolor{BrickRed}{energy storage system} to enhance the voltage stability of power systems with a high penetration of \textcolor{BrickRed}{renewable energy sources}.



\appendices

\section{Sample Path Concentration for SDE Model}
\label{appendix_samplepath}
We consider the deterministic slow-fast system (\ref{general slow fast}),
in which $\mathcal{M}_0$ and $\mathcal{M}_{\epsilon}$ are the slow and the invariant manifolds respectively, as defined in Section \ref{slowfastsystem_subsection}, and the stochastic slow-fast system (\ref{general slow fast SDE_appendix}).
\begin{eqnarray}
    \dot{\bm{x}}&=&\frac{1}{\epsilon}\bm{f(x,y)}+\frac{\sigma}{\sqrt{\epsilon}}\bm{f_1(x,y)}\bm{\xi}, \quad \bm{x}\in \mathbb{R}^{n_x}\nonumber\\
    \dot{\bm{y}}&=&\bm{g(x,y)}+\textcolor{BrickRed}{\tilde{\sigma}}\bm{g_1(x,y)}\bm{\xi},\qquad \bm{y}\in \mathbb{R}^{n_y} \label{general slow fast SDE_appendix}
\end{eqnarray}
where $\int_{0}^{t} \bm{\xi}(u) du$ is a $k-$dimensional Brownian motion and $\bm{f_1} \in \mathbb{R}^{n_x \times k}, \bm{g_1} \in \mathbb{R}^{n_y \times k}$ are sufficiently smooth functions. For the noise intensities $\sigma$ and $\textcolor{BrickRed}{\tilde{\sigma}}$, we focus on the case where $\textcolor{BrickRed}{\tilde{\sigma}}$ does not dominate $\sigma$, i.e.,
$\textcolor{BrickRed}{\tilde{\sigma}}=\rho\sigma$ where $\rho$ is uniformly bounded above in $\epsilon$.

\textit{Assumption 1.}
\begin{itemize}
    \item $\bm{f}$, $\bm{g}$, $\bm{f_{1}}$ and $\bm{g_{1}}$ and all their partial derivatives up to order 2 (respectively 1) are uniformly bounded in norm in $D$,  $\bm{f} \in C^2(D, \mathbb{R}^{n_x})$, $\bm{g} \in C^2(D, \mathbb{R}^{n_y})$, $\bm{f_{1}} \in C^1(D, \mathbb{R}^{n_x \times k})$,  $\bm{g_{1}} \in C^1(D, \mathbb{R}^{n_y \times k})$      where $D$ is an open subset $D \subset \mathbb{R}^{n_x}\times\mathbb{R}^{n_y}$ and $n_x, n_y, k \geq 1$.
    \item The slow manifold $\mathcal{M}_0$ of the deterministic system is a stable component of the constraint manifold.
    \item Matrix $\bm{f_{1}}(\bm{x}, \bm{y})\bm{f_{1}}(\bm{x}, \bm{y})^T$ is positive definite for $\bm{y} \in D_{\bm{y}}$  $\subset \mathbb{R}^{n_y}$.
\end{itemize}

According to Theorem 5.1.6 in \cite{Gentz06} (see \textit{Theorem 4} below), with a probability roughly like $1-O(e^{-h^2/{2\sigma^2}})$, the sample paths of (\ref{general slow fast SDE_appendix}) remain concentrated up to time $t$ in the ellipsoidal layer that surrounds the invariant manifold and is defined as:
\begin{equation}
{\cal B}(h) = \{( \bm{x},\bm{y}):\langle \bm{x}-\bm{\bar{x}}(\bm{y},\epsilon),{\bm{\bar{X}}}(\bm{y},\epsilon)^{-1}( \bm{x-\bar{\bm{x}}}(\bm{y}, \epsilon)) \rangle<h^2 \}
\label{neighborhood}
\end{equation}
where $\bm{\bar{X}}(\bm{y},\epsilon)$ describes the cross section of ${\cal B}(h)$  and is well defined (See Chapter 5.1.1 of \cite{Gentz06} for more details).

\textit{Theorem 4} \cite{Gentz06}:  If \textit{Assumption 1} holds, there exist $\epsilon_{0}>0, h_{0}>0, \delta_{0}>0$ and a time $t_{1}$ of order $\epsilon |\log h|$ such that, for $\delta \leq \delta_{0}$, if the initial condition $(\bm{x}(0),\bm{y}(0))$ where $\bm{y}_{0} \in D_{\bm{y}}$ satisfies $ (\bm{x}(0),\bm{y}(0)) \in {\cal B}(\delta)$:

\begin{eqnarray}
   && \mathbb{P}\{\exists s\in[t_1,t]: (\bm{x},\bm{y})\not\in{\cal B}(h)\}\nonumber\\
   &&\leq C_{n_{\bm{x}},n_{\bm{y}}}(t,\epsilon) e^{\frac{-h^2}{2\sigma^2}(1-O(h)-O(\epsilon))}\label{eq:probabilityofleaving_appendix}
\end{eqnarray}
for all $t \geq t_{1}, \epsilon\leq \epsilon_0$, $h\leq h_0$, where the coefficient  $C_{n_{\bm{x}},n_{\bm{y}}}(t,\epsilon)=[C^{n_{\bm{y}}}+h^{-n_{\bm{x}}}](1+\frac{t}{\epsilon^2})$ is linear in time.

\section{Cross Section of $\cal {N}$$(h)$}
\label{appendix_crosssection}



The cross section $\bm{\bar{U}}(\bm{p},\epsilon t)$ of $\cal {N}$$(h)$ is the solution of the following slow-fast system:

\begin{eqnarray}
    \epsilon U^{\prime}&=&A(\bm{p}, y)U+UA(\bm{p}, y)^T + \begin{bmatrix} \kappa I_{n_{\bm{x}}} & \bm{0} \\ \bm{0} & B_{\bm{\eta}}B_{\bm{\eta}}^T   \end{bmatrix}  \nonumber \\
    y^{\prime} &=&1
    \label{cross section system}
\end{eqnarray}
where $\kappa>0$ and sufficiently small to ensure positive definiteness and
\begin{eqnarray}
     A(\bm{p},y) &=& \partial_{\bm{u}} \bm{G}(\bm{u^{\star}}(\bm{p},y),\bm{p},y)\\
        y&=&\epsilon t \nonumber
\end{eqnarray}

Hence, $\bm{\bar{U}}(\bm{p},\epsilon t)$ is well defined.


\bibliography{revisedjournal}
\vspace{-20pt}
\begin{IEEEbiography}[{\includegraphics[width=1in,height=1.25in,clip,keepaspectratio]{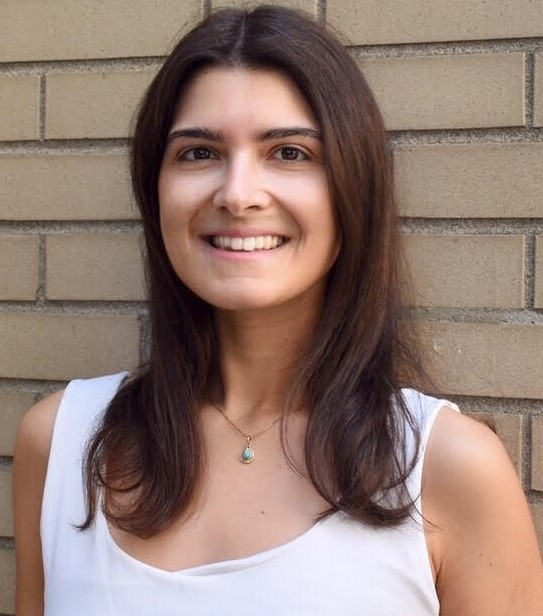}}]%
{Georgia Pierrou}
received a Diploma in Electrical and Computer Engineering from the National Technical University of Athens, Athens, Greece in 2017. Since 2017, she has been
pursuing the Ph.D. degree in Electrical Engineering with the Electric Energy Systems Laboratory at McGill University, Montreal, Canada. Her research interests include power system dynamics, control and uncertainty quantification.
\end{IEEEbiography}

\begin{IEEEbiography}[{\includegraphics[width=1in,height=1.25in,clip,keepaspectratio]{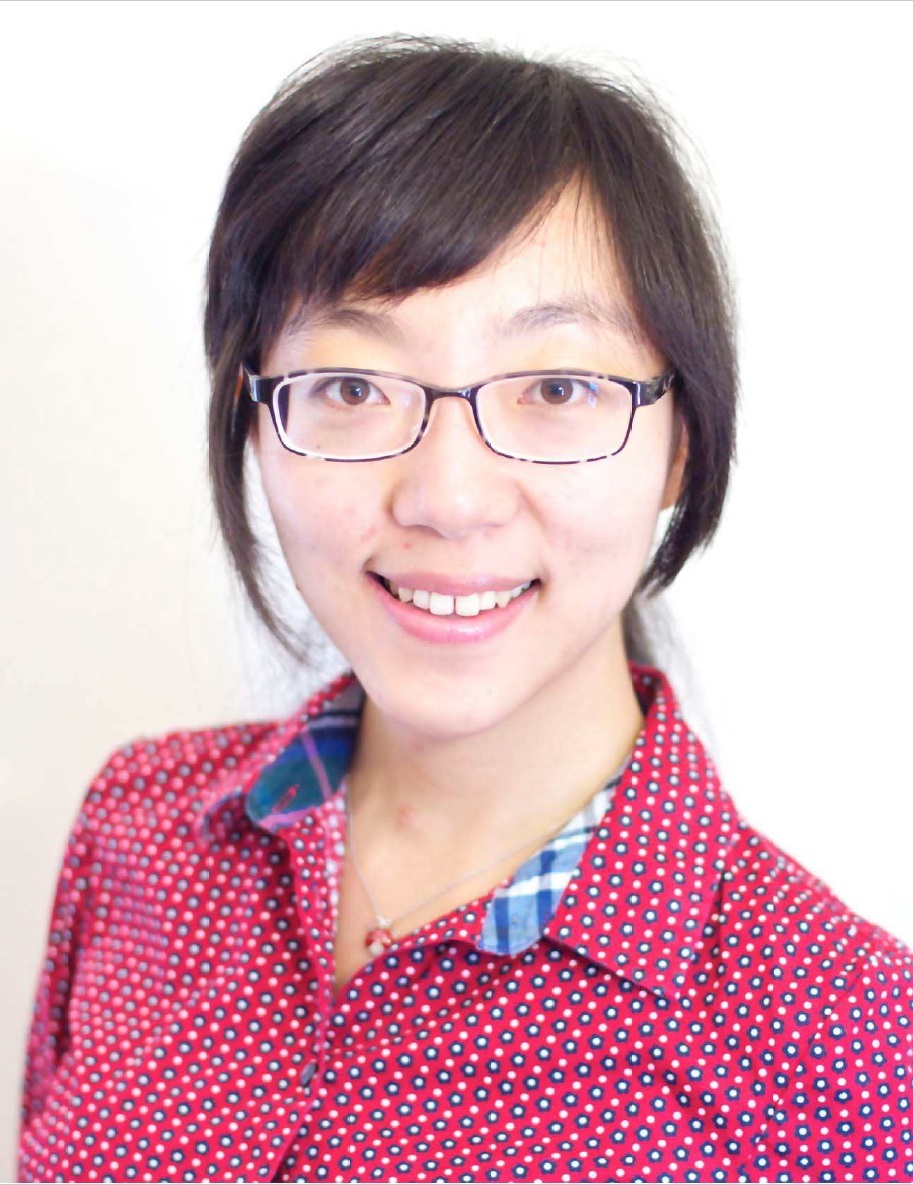}}]%
{Xiaozhe Wang}
is currently an Assistant Professor in the Department of Electrical and Computer Engineering at McGill University, Montreal, QC, Canada. She received the Ph.D. degree in the School of Electrical and Computer Engineering from Cornell University, Ithaca, NY, USA, in 2015, and the B.S. degree in Information Science \& Electronic Engineering from Zhejiang University, Zhejiang, China, in 2010. Her research interests are in the general areas of power system stability and control, uncertainty quantification in power system security and stability, and wide-area measurement system (WAMS)-based detection, estimation, and control. She is serving on the editorial boards of IEEE Transactions on Power Systems, Power Engineering Letters, and IET Generation, Transmission and Distribution.
\end{IEEEbiography}

\end{document} 
\n